\documentclass[10pt]{article}
\def\be{\begin{equation}}
\def\ee{\end{equation}}

\def\bea{\begin{eqnarray}}
\def\eea{\end{eqnarray}}

\begin{document}

\hspace{1.0cm} \parbox{15.0cm}{

\baselineskip = 15pt

\noindent {\bf MORPHOLOGY OF CIRCUMSTELLAR ENVIRONMENT AND \\
SOME CHARACTERISTICS OF CIRCUMSTELLAR SHELLS OF STARS \\
WITH THE R CORONAE BOREALIS VARIABILITY}

\bigskip
\bigskip

\noindent {\bf A.E.Rosenbush}

\bigskip

\baselineskip = 9.5pt

\noindent {\small \copyright~2000}

\smallskip

\noindent {\small {\it Main Astronomical Observatory of
National Academy of Sciences of Ukraine\\
03680, Golosiiv, Kyiv-127, Ukraine}} \\
\noindent {\small {\it e-mail:}} {\tt mijush@mao.kiev.ua}

\baselineskip = 9.5pt \medskip

\medskip \hrule \medskip

\noindent The well-known light minima of stars with the
R Coronae Borealis variability are caused by the formation of
an additional circumstellar dust shell, the screening shell,
inside the permanent shell.

Under the assumption of uniform distribution of matter in the
circumstellar environment we estimated the optical thickness of
the permanent gas--and--dust shell at 0.2--0.7, and its geometrical
thickness is no less than 0.4 of its own radius. The wavelength
dependence of extinction is close to neutral.

From spectral observations of R CrB itself in the 1985 minimum
we traced the transformation of the stellar linear and molecular
absorption spectrum to the emission spectrum and established that
the fast variation of the U--B colour index by --0.6 in the light
decline was caused purely by a change of the spectrum type. The
spectrum transformation causes an increase of star brightness
in the U, B, and V bands by about 1.4, 0.75, and 0.75 mags,
correspondingly.

It is suggested that a high--velocity ($>$200 km/s) matter
stream through the circumstellar environment is the cause of
the excitation of the emissions observed during light minima
when the photospheric flux is weakening.

\medskip \hrule \medskip

}
\vspace{1.0cm}

\baselineskip = 11.2pt

\noindent {\small {\bf INTRODUCTION}}
\medskip

\noindent The set of unique characteristics of the stars with the
R Coronae Borealis (RCB) type variability keeps constantly the high
interest to them from observers and theorists. For all this, side
by side with the progress in observations, negligibility in the theory
is noted.

The principal variability -- deep light declines up to 8 mag which
last hundreds of days and are caused by dust condensation on the
line of sight -- remains the phenomenon with many questions,
including even the lack of a simple geometric model of the
phenomenon \cite{R96a,F96,C96,R96b}.

The basis of the modern model is the cloud structure of the
circumstellar permanent dust shell \cite{F96,C96} when one of
the dust clouds in this shell forms on the line of sight, and,
as a result, a visual light minimum is observed. This explains
the lack of anticorrelation between variations of visual and infrared
brightness during a minimum \cite{FGS72}. The lack of clear influence of
the permanent shell, in which up to 40 percent of the bolometric
luminosity of star is transformed, on apparent characteristics of
the star at a light maximum supports the idea of its cloud structure.
The difficulties of this model in simulating light minima were well
known \cite{C96,R96b}.

As the result of our studies, we arrived at the understanding that
it is necessary to investigate not variables of the R Coronae
Borealis type but the R Coronae Borealis phenomenon observed in
various objects, for instance, in novae. It is need the high mass
loss rate and the enhanced abundance of carbon. The hydrogen is
necessary for the phenomenon. Hydrogen deficiency is a consequence
of the phenomenon, and its full exhaustion means the cessation of
the variability. This view helps us to understand many sides of the
phenomenon. At a time we concluded that the 1972 rejection of the
model of a spherical dust shell for the interpretation of the R Coronae
Borealis phenomenon \cite{FGS72} was not sufficiently justified:
not all sides of the phenomenon were known and were taken into
consideration. Below we consider some principal consequences from
the assumption of the uniform distribution of matter in the
circumstellar environs.

\bigskip
\noindent {\small {\bf STRUCTURE OF CIRCUMSTELLAR ENVIRONS }}
\medskip

\noindent Three spherical shells can exist in the environs of a star:
two constant shells, the permanent and the fossil (the latter is
not connected directly with current light minima), and a temporal,
screening one. Both shells are formed in the outflow of mass with
a high loss rate estimated at about $10^{-6}$ mass of the Sun per year.

The radius of permanent shell was estimated at 26 radii of the
star or 2340 radii of the Sun \cite{R96a}. The formation of this
shell can occur in the framework of the homogeneous dust nucleation
theory by Fadeyev \cite{F88}, which gives the dust condensation zone
radius close to this value. The mass of the shell is about $10^{-6}$
mass of the Sun.

Sometimes conditions can arise for dust condensation very close
to the star, i.e., a screening shell with a radius of 4-10 radii
of star is formed \cite{R96a}. We estimated the full mass of this
shell at about $10^{-7}$ mass of the Sun. The screening shell has
its own infrared excess which compensates the weakening of the
permanent shell luminosity. The effective temperature of screening
shell during its maximum optical thickness drops down to 700-900 K,
which is equal to the permanent shell temperature.

\bigskip
\noindent {\small {\bf OPTICAL THICKNESS OF PERMANENT SHELL }}
\medskip

\noindent It is known that the permanent shell re--radiates up to 42
percent of the bolometric luminosity of the star itself \cite{FGS72}.
If we assume that the shell only re--radiates, we have an indirect
estimate of optical thickness of this shell at about 0.7.

FG Sge allows a more direct estimation. Since 1992 it shows
the RCB type variability that allows us to compare stellar
parameters before and after 1992. Since 1992 FG Sge was at the state
of maximal brightness, i.e., it had only the permanent shell,
in the end of 1997 - the beginning of 1998, in 1999 and 2000. The
difference of average brightness in these periods from the level
before 1992 may be due to the extinction in the permanent
shell. Our estimate for this is about 0.26$^m$. It means that
the permanent shell absorbs nearly 20 percent of the bolometric
luminosity of the star itself. From published data we can estimate
that the bolometric luminosity of shell is nearly 25 percent of
the bolometric luminosity of the star itself.

\bigskip
\noindent {\small {\bf SIZE OF DUST PARTICLES }}
\medskip

\noindent RY Sgr has a number of observations in the UV, where
dust particles primarily show an evidence of their existence after
condensation.

From the graphic and tabulated data of \cite{CWS92}, in addition to
the author's results, we can obtain the following data.

Firstly, the light minima were not seen in the UV. The decline of the
240 nm flux during a minimum occurs probably only up to the
weakening of visual brightness by nearly 2$^m$. The following rise of
extinction in the visual is not accompainied by a rise of extinction
in the UV. According to \cite{DL84}, this may occur by the change
of a dust particle size from 0.01 micron to 0.3 micron. The particle
density does not increase, otherwise the extinction should continue
to increase. From this we can draw the unexpected and interesting
conclusion that the light minimum may be considered only as a
consequence of a change of the dust particle sizes but not their
density, and the dust nuclei are constantly present in the stellar
atmosphere.

Secondly, the amplitude of the UV brightness pulsations outside minima
is a factor of 1.8 higher, on the average, than in the visual
pulsations. The higher amplitude of the UV pulsations can be
interpreted as the consequence of condensation of dust particles,
their growth up to radii of 0.003 micron and disruption in every
pulsation.

The dust particles move away from the star, dragging the gas of
the stellar atmosphere.

\bigskip
\noindent {\small {\bf OUTFLOW VELOCITY AND  ORIGIN  OF
EMISSION LINES }}
\medskip

\noindent The kinetic energy of helium atoms which are the main gas
component of the stellar atmosphere is as high as 800 eV and more
at a 200 km/s velocity. This energy is sufficient to excite
emission by the interaction with low--velocity atoms. It is
known that emissions of C IV] 155.0 nm, C II 133.5 nm, etc. are
observed beyond minima, i.e., we have a clear indication on their
origin in the permanent shell. If there is no dust, there are no
UV emissions mentioned above. This is observed in XX Cam and
HD 182040.

This transformation of energy occurs on the inner boundary of
permanent shell on a level of 10--26 star radii at a density
of about $10^{-14}$--$10^{-15}$ g/cm or a helium density of about
$10^{9}$ cm$^{-3}$.

Thus we can identify the region of the origin of sharp emission
lines or the "chromosphere" of the star with the zone of dust
condensation or with the permanent and screening dust shells.

During light minima the absorption spectrum disappears and the
emission spectrum appears. Such transformation determines almost
completely the behaviour of U--B and B--V indices observed during
a visual minimum.

\bigskip
\noindent {\small {\bf VARIATIONS OF COLOUR INDICES AND
SPECTRUM IN A LIGHT MINIMUM }}
\medskip

\noindent Based on the R CrB minimum in 1985, for which our
extensive spectral observations are available and a detailed
photometry is published, we studied the effect of the disappearance
of absorption lines and appearance of emission lines on the colour
indices of the broad--band photometric U, B, and V system.

We obtained that the key points of spectral changes are three
levels of brightness weakening.

I   -- 1.4--1.6$^m$  -- no considerable changes;

II  -- 2.9--3.4$^m$  -- the disappearance of absorptions and the
appearance of emissions;

III -- 4.3$^m$   -- the maximum of emission intensity and its following
sharp drop.

It is necessary to compare these levels with the light and colour
curves. After the light fading to the first level a sharp decrease
of U--B began. At the second level, a sharp increase of B--V occurred.
At the third level, the brightness is minimal, it is the reverse point
of the light curve and a sharp extremum of U--B; the B--V behaviour
did not change.

Thus we make the opposite conclusion that the colour change is
due mainly to the change from the absorption spectrum to the
emission, and all variations in U--B from +0.3 to --0.3$^m$ in
1985 were related to this cause only. The B--V variation did not
exceed --0.1$^m$. The estimation of spectral transformation on our
spectrograms for the 1985 minimum in the B band gives a value of
line contribution of about 0.75$^m$. Thus one can say that the lines
contribute about 0.75$^m$ in the B band, and their contribution
in the V band was slightly smaller. The contribution in the U band was
well above, and it reached 1.4$^m$. The light minima are distinguished
by the intensity of emission lines. Therefore, the colour variations
were dissimilar but the cause of variations is the same.

This moves away one of the principal objections against the
uniform distribution of dust in the circumstellar shells.

The well--known loops of a star on the "colour--brightness"
diagram during its light minimum should be interpreted with
this circumstance taken into consideration.

\bigskip
\noindent {\small {\bf CONCLUSION }}
\medskip

\noindent As an addition to the above, we present a possible
version of the development of the RCB type variability in
FG Sge as a consequence of formation of an uniform circumstellar
dust shell.

Years before 1992 FG Sge had a small infrared excess
appropriate to of a low dust mass of 4*$10^{-10}$ mass of the Sun.
The IR excess was constantly rising, and at some stage of its
increase the first RCB minimum occurred. This behaviour is very
similar to the RCB phenomenon in the DQ Her novae \cite{R96a}, when
the IR excess appeared well before the beginning of the RCB--like
visual decline at the transition stage of a nova outburst. I.e.,
in FG Sge the first RCB minimum was also connected with the
formation of a screening shell. At the same time there was
no permanent gas--and--dust shell, as no characteristic emissions
were observed during the first minimum, but only the blue-shifted
circumstellar absorption D Na I lines. It is unlikely that the
first dust cloud as a cause of a light fading is formed at once on
the line of sight. If this were so, the "chromospheric" emissions
would necessarily be observed in the 1992 minimum as they were
observed in following ones. This supports our assumption about
the origin of emissions in the permanent shell.

Since 1972 we have again all grounds to use the dust shell model
to interpret the R Coronae Borealis phenomenon. The lack of
obvious progress in the theory is connected with the one--sided
approach to this "unique" variability. It is a widespread phenomenon,
and it is more correct to study the RCB phenomenon, observing
various objects, including novae at the transition stage of outburst.

The full text will be published as two papers in the
"Astrofizika/Astrophysics" journal in 2000 and 2001.

\bigskip
\noindent {\small {\bf REFERENCES}}


\end{document}